\documentclass[aps, prl, amsmath, floats, floatfix, twocolumn,superscriptaddress, nofootinbib, showpacs,reprint]{revtex4}

\usepackage{graphicx}
\usepackage{amsmath,amssymb}
\usepackage{amsfonts}
\usepackage{xspace} 
\usepackage[usenames]{color}
\usepackage{dcolumn}
\usepackage{bm}
\usepackage{mathrsfs}
\usepackage[colorlinks=true]{hyperref}
\usepackage[all]{hypcap} 
 \usepackage[utf8]{inputenc} 

\usepackage{ulem}
\definecolor {darkgreen}{rgb}{0.2,0.7,0.2}

\newcommand{\eq}{\begin{equation}}
\newcommand{\be}{\begin{equation}}
\newcommand{\eeq}{\end{equation}}
\newcommand{\ee}{\end{equation}}

\begin{document}

\title{Gravitational-wave emission in shift-symmetric Horndeski theories}

\author{Enrico Barausse}
\affiliation{Sorbonne Universit\'es, UPMC Univ Paris 6 \& CNRS, UMR 7095, Institut d’Astrophysique de Paris, 98 bis bd Arago, 75014 Paris, France}
%\affiliation{CNRS, UMR 7095, Institut d'Astrophysique de Paris, 98bis Bd Arago, 75014 Paris, France}
%\affiliation{Sorbonne Universit\'es, UPMC Univ Paris 06, UMR 7095, 98bis Bd Arago, 75014 Paris, France}
\author{Kent Yagi}
\affiliation{Department of Physics, Princeton University, Princeton, NJ 08544, USA.}
\affiliation{Department of Physics, Montana State University, Bozeman, MT 59717, USA.}

\begin{abstract}
Gravity theories beyond General Relativity typically predict dipolar gravitational emission by compact-star binaries. This emission is sourced by ``sensitivity'' parameters depending on the stellar compactness. We introduce a general formalism to calculate these parameters, and show that in shift-symmetric Horndeski theories stellar sensitivities and dipolar radiation vanish, provided that the binary's dynamics is perturbative (i.e.~the post-Newtonian formalism is applicable) and cosmological-expansion effects can be neglected. This allows reproducing the binary-pulsar observed orbital decay.
\end{abstract}

\date{\today \hspace{0.2truecm}}

\maketitle

General Relativity (GR) is very successful at interpreting gravity on a huge range of scales, field strengths and velocities. 
Nevertheless, evidence for Dark Matter and Dark Energy  may be interpreted as a breakdown of GR on cosmological scales. Also, GR is intrinsically incompatible with quantum field theory, and should be replaced, at high energies, by a (still unknown) quantum theory of gravity. Given this situation, guidance may come from experiments, namely those testing gravity in regimes involving strong fields and/or relativistic speeds. These experiments include measurements of the gravitational-wave (GW) driven orbital decay of binary pulsars~\cite{damour-taylor}, and upcoming GW interferometers~\cite{ligo,virgo,kagra}. It is therefore crucial to analyze gravitational emission in theories alternative to GR.

Modified gravity theories typically generalize GR by introducing extra gravitational fields non-minimally coupled to the metric.  
An example is Fierz-Jordan-Brans-Dicke (FJBD) gravity~\cite{Fierz:1956zz,Jordan:1959eg,Brans:1961sx}, see e.g.~Ref.~\cite{Berti:2015itd} for a recent review of more theories.
Often, even theories where no extra fields are explicitly added [e.g.~$f(R)$ gravity]
can be recast as 
GR plus extra fields by a suitable change of
variables. The extra fields generally introduce ``fifth forces'', and the motion of a body free-falling
in a gravitational field will generally depend on the body's nature. This effect can be suppressed 
for weakly gravitating bodies by assuming that the extra gravitational fields do not couple to matter directly, i.e.~the equivalence principle (EP) can be
restored for weakly gravitating bodies (``weak EP''). However, for bodies with strong self-gravity, the extra fields will still effectively
couple to matter (because they are non-minimally coupled to the metric, which in turn is coupled to matter via gravity). This coupling will be increasingly important
as the body's gravitational binding energy -- which measures the ``strength'' of the body's self-gravity -- increases. Indeed, the extra gravitational fields will generally
affect the body's binding energy, and since in relativistic theories all forms of energy gravitate, the body's gravitational mass will depend on the extra fields via the binding energy. 
As such, the inertial mass may differ from the gravitational mass if the binding energy's contribution to the latter is important. Indeed, possible deviations
from the ``strong'' EP (i.e.~the universality of free-fall for strongly gravitating bodies) in modified gravity theories are typically parametrized by the ``sensitivities''~\cite{eardley} 
\begin{equation}\label{eq:sens}
s_{_{Q_A}}= \frac{1}{M}\frac{\partial  M}{\partial Q_A}\Bigg\vert_{N,\Sigma}\,,
\end{equation}
i.e.~the derivatives of the gravitational mass $M$ relative to the theory's extra fields $Q_A$,
while keeping the body's total baryon number $N$ and entropy $\Sigma$ fixed. (The sensitivities thus measure the body's response to changes in the local value of $Q_A$.)
For weakly gravitating bodies to obey the weak EP, it must be $s_{_{Q_A}}\approx 0$. Indeed, 
if the extra fields  $Q_A$ do not couple to matter directly, they only enter $M$ via the binding energy, whose contribution
is negligible if the body's self-gravity is weak. 

The sensitivities enter both the conservative dynamics of binary systems (e.g.~the periastron precession) and the dissipative one (i.e.~GW emission).
Their leading-order dissipative effect is the emission of dipolar gravitational radiation, i.e.~binaries will produce GWs
$\phi \sim G/c^3\times {\cal O}[(s^{_{(1)}}_{\phi}-s^{_{(2)}}_{\phi})^2]$, where $s^{_{(1)}}_{\phi},\,s^{_{(2)}}_{\phi}$ are the bodies' sensitivities. This is a $-1$ post-Newtonian (PN) effect, i.e.~it is enhanced by $(v/c)^{-2}$ compared to GR ($v$ being the binary's relative velocity), if $s^{_{(1)}}_{\phi}$, $s^{_{(2)}}_{\phi}\sim 1$. 
Therefore, knowledge of the sensitivities is crucial to verify the agreement between a theory and GW observations (binary-pulsar data or direct detections). 
For example, binary pulsars have already placed strong constraints on 
Lorentz-symmetry violations in gravity~\cite{PhysRevD.89.084067,Yagi:2013qpa} and on certain scalar-tensor theories~\cite{freire,Wex:2014nva,Berti:2015itd}, and even stronger bounds will be possible with direct GW detections~\cite{ST1,ST2,Sampson:2014qqa}.

Here, we generalize previous work in scalar-tensor theories~\cite{damour_esposito_farese,ST0} and Lorentz-violating gravity~\cite{PhysRevD.89.084067,Yagi:2013qpa} by introducing a formalism 
 to calculate the sensitivities of stars (including pulsars) in \emph{generic} theories. As an application, we calculate them in
 the most general scalar-tensor theories (``Horndeski theories'' or ``generalized galileons'')~\cite{Horndeski:1974wa,Deffayet:2009mn,Deffayet:2011gz} 
that have second-order field equations and are invariant under a shift of the scalar field $\phi$, i.e.~$\phi\to\phi\ +$ constant.
These theories have received much attention because \textit{some} of them provide a screening mechanism (``Vainshtein mechanism''~\cite{Vainshtein:1972sx,Deffayet:2001uk}; see also Refs.~\cite{Babichev:2013usa,Winther:2015pta}), which may permit modifying gravity on cosmological scales (possibly reproducing cosmological data without Dark Energy) while
recovering GR on small scales, where gravitational modifications would be ``screened''.
Also, galileon interactions arise in the decoupling limit of massive gravity~\cite{deRham:2010ik,deRham:2010kj}.

We show that in shift-symmetric Horndeski theories (SSHTs) stellar sensitivities vanish\footnote{This agrees with previous
results for a specific theory of this class, i.e.~Einstein-dilaton Gauss-Bonnet gravity in the decoupling limit~\cite{Yagi:2011xp}.}, 
and  the leading-order GW emission
matches GR's. We conclude that SSHTs reproduce existing binary-pulsar data (provided that 
a PN expansion over Minkowski space is adequate -- i.e.~the binary's dynamics is perturbative --, and that 
cosmological-expansion effects can be neglected~\cite{Jimenez:2015bwa,Babichev:2012re,Babichev:2013cya,Jacobson:1999vr}), but deviations might still appear for the sources targeted by upcoming GW interferometers.

We use Latin (Greek) lower-case letters for space (spacetime) indices, and capital Latin letters for indices running on fields. Repeated indices denote summations, %overdots partial time derivatives, 
and we assume $c=1$ and signature $(-,+,+,+)$.

%%%%%%%%%%%%%%%%%%%%%%%%%%%%%
\emph{A general expression for the sensitivities}:
Consider an action 
\begin{equation}\label{generic_action}
S=\int {\cal L}(Q_A,\partial_\mu Q_A) d^4 x,
\end{equation}
where $Q_A$ are fields,
%The Hamiltonian is
%\begin{equation}
%H=\int d^3 x \left( \pi_A \dot{Q}_A- {\cal L}\right),
%\end{equation}
%where 
%$\pi_A={\partial {\cal L}}/{\partial \dot{Q}_A}$
%are the momenta conjugate to $\dot{Q}_A\equiv\partial_t Q_A$.
%
and the corresponding field equations
\begin{equation}\label{ELeq}
\partial_\mu \left(\frac{\partial \cal{L}}{\partial (\partial_\mu Q_A)}\right)-\frac{\partial \cal{L}}{\partial Q_A}=0\,.
\end{equation}
A solution's canonical mass-energy is
\begin{equation}
M=\int d^3x \left(\pi_A\partial_t{Q}_A -{\cal L}\right)\,,
\end{equation}
where $\pi_A={\partial {\cal L}}/{\partial (\partial_t{Q}_A)}$ for brevity.
For a solution with $\pi_A\partial_t{Q}_A=0$ (this is the case e.g.~for stationary solutions, where $\partial_t{Q}_A=0$, or solutions 
where some variables $Q_A$ depend on time, but
 $\pi_A=0$, see e.g.~Eq.~\eqref{fluid_action}),
the mass is then simply\footnote{We have checked that this mass matches the ADM mass, both in GR and SSHTs (see also Ref.~\cite{Iyer:1994ys}).}
\begin{equation}\label{canonical mass}
M =-\int d^3 x\, {\cal L}(Q_A,\partial_\mu Q_A)\,.
\end{equation}
Consider a neighboring solution, with $\pi_A\partial_t{Q}_A=0$ and 
 the same total baryon number and entropy [c.f. Eq.~\eqref{eq:sens}]. The two solutions represent the same star with different local values of 
the fields $Q_A$, and their mass
difference is
\begin{equation}\label{dM}
\delta M =
- \int d^3 x \partial_t \left(\pi_A \delta Q_A\right)
-\int d^2 S_i \frac{\partial {\cal L}}{\partial (\partial_i Q_A)}\delta Q_A\,,
\end{equation}
where $\delta Q_A$ is the difference between the fields, 
$d^2 S_i$ is a coordinate surface element, 
and we have used  Eq.~\eqref{ELeq} to show that the bulk terms vanish, as well as Gauss's theorem.

Because this result assumes an action with no derivatives higher than first order, it would not seem to apply to GR, since the  Einstein-Hilbert action depends on second metric derivatives. 
However, the latter enter the Einstein-Hilbert action only through a total divergence, i.e.~GR can be described by a first-order ``Einstein Lagrangian'' ${\cal L}_{g}=\sqrt{-g} g^{\mu\nu}
  (\Gamma^\alpha_{\mu\lambda}\Gamma^{\lambda}_{\nu\alpha}-\Gamma^\lambda_{\mu\nu}\Gamma^\alpha_{\lambda\alpha})/(16 \pi G)$
  (see e.g.~Ref.~\cite{landau}).  
Besides the metric, in theories different from GR there are other gravitational degrees of freedom, which we denote by $\phi_A$, and which we assume coupled to the metric and its derivatives, i.e.~with Lagrangian ${\cal L}_\phi(\phi_A, \partial_\mu \phi_A, g_{\mu\nu}, \partial_\alpha g_{\mu\nu})$.
Moreover, the matter fields $\psi_B$ must couple minimally to the metric [i.e.~with Lagrangian ${\cal L}_m(\psi_B,\partial_\mu\psi_B,g_{\mu\nu})$] to satisfy the weak EP\footnote{The fields $\phi_A$ and $\psi_B$ are not necessarily scalars, e.g.~they could represent the components of a vector or tensor.}.
Consider a stationary (i.e.~time independent) star in one such theory. The mass difference between neighboring solutions is
\begin{multline}\label{deltaM}
\delta M = -\int d^2 S_i \frac{\partial {\cal L}_g}{\partial (\partial_i g_{\mu\nu})}\delta g_{\mu\nu}\\-\int d^3 x \partial_t \left(\pi_{\psi_B} \delta \psi_B\right)-\int d^2 S_i \frac{\partial {\cal L}_m}{\partial (\partial_i \psi_B)}\delta \psi_B\\- \int d^2 S_i \frac{\partial {\cal L}_\phi}{\partial (\partial_i \phi_A)}\delta \phi_A-\int d^2 S_i \frac{\partial {\cal L}_{\phi}}{\partial (\partial_i g_{\mu\nu})}\delta g_{\mu\nu}\,,
\end{multline}
where we have used $\partial_t ( \pi_{g_{\mu\nu}}\delta g_{\mu\nu})=\partial_t (\pi_{\phi_A}\delta \phi_A)=0$ because of stationarity. Note that we have {\it not} 
assumed $\partial_t (\pi_{\psi_B} \delta \psi_B)=0$ [but only $\pi_{\psi_B} \partial_t \psi_B=0$,
so that Eqs.~\eqref{canonical mass} and \eqref{dM} hold], which will allow using this expression for perfect-fluid stars. 

The Lagrangian for a perfect fluid with equation of state $\rho=\rho(n,\sigma)$ 
($\rho$, $n$ and $\sigma$ being respectively the energy density, baryon-number density and entropy per particle) is~\cite{PhysRevD.2.2762,Brown:1992kc} 
\begin{equation} \label{fluid_action}
{\cal L}_m=-\sqrt{-g} \rho\\-\varphi \partial_\mu J^\mu- \theta  \partial_\mu (\sigma J^\mu)-\alpha_A \partial_\mu(\beta_A J^\mu)\,,
\end{equation}
where $\alpha_A$, $\beta_A$, $\varphi$ and $\theta$ are scalars ($\alpha_A$ can be interpreted as Lagrangian coordinates).
The fluid four-velocity is defined as $U^\mu=J^\mu/|J|$ (with $|J|=\sqrt{-g_{\mu\nu} J^\mu J^\nu}$)
and  $n=|J|/\sqrt{-g}$, i.e.~$J^\mu$ is the
baryon number density current $J^\mu = \sqrt{-g} n U^\mu$. Variation with respect to $g_{\mu\nu}$ (keeping $J^\mu,\varphi,\theta,\sigma,\alpha_A,\beta_A$ fixed)
gives the perfect-fluid stress-energy tensor, by using the first law of thermodynamics
$n\partial \rho/\partial n\vert_{\sigma}= p+\rho$, ($p$ being the pressure). Variations with respect to $J^\mu,\varphi,\theta,\sigma,\alpha_A,\beta_A$
yield 
\begin{multline}\label{fluid_eqs}
\partial_\mu J^\mu=\partial_\mu (\sigma J^\mu)=J^\mu\partial_\mu\beta_A=J^\mu\partial_\mu\alpha_A=0\,,\\
h U_\mu=-\partial_\mu\varphi-\sigma \partial_\mu \theta-\beta^A\partial_\mu \alpha_A\,,\quad
U^\mu \partial_\mu \theta= T\,,
\end{multline}
where $h=(p+\rho)/n$ is the specific enthalpy, and 
we have used ${\partial \rho}/{\partial \sigma}\vert_n = nT$ ($T$ being the temperature) from the first law of thermodynamics. From these equations, it is
clear that $\varphi$ and $\theta$ are Lagrange multipliers enforcing the local baryon-number and entropy conservation, while $\beta_A$
and $\alpha_A$ are constant along the fluid lines. Also, these equations 
imply the conservation of the fluid stress-energy tensor~\cite{PhysRevD.2.2762,Brown:1992kc}.%, i.e.~the relativistic energy-conservation and Euler equations. 

For a stationary fluid, we can adopt comoving coordinates where $U^i=J^i=0$. Therefore,
the third term in Eq.~\eqref{deltaM} vanishes (since $\delta J^i=0$), while the second becomes
\begin{multline}\label{fluid1}
- \int d^3 x \partial_t \left(\pi_{\psi_B} \delta \psi_B\right)\\=\int d^3 x \partial_t \left[\varphi \delta J^t+\theta \delta (\sigma J^t)+ \alpha_A \delta (\beta^A J^t) \right]
\\=-\int d^3 x \left[(h-\sigma T) U_t\delta J^t+T U_t \delta (\sigma J^t) \right]\,,
\end{multline}
where we have used
$\partial_t\alpha_A=\partial_t\beta^A=\partial_t J^t=\partial_t \sigma=0$, 
$-\partial_t\varphi= (h -\sigma T) U_t$ and $-\partial_t\theta= T U_t$,
obtained from Eq.~\eqref{fluid_eqs} (with $J^i=0$).
For a fluid
in hydrostatic and thermodynamic equilibrium in a stationary spacetime, $T U_t$ and $(h-\sigma T) U_t$ are uniform~\cite{MTW,rezzolla_book,eric}, and Eq.~\eqref{fluid1} thus becomes
\begin{equation}
- \int d^3 x \partial_t \left(\pi_{\psi_B} \delta \psi_B\right)= -(h-\sigma T) U_t\delta N-T U_t \delta \Sigma\,,\end{equation}
where $\delta N=\int d^3 x \delta J^t$ and $\delta \Sigma= \int d^3 x \delta (\sigma J^t)$ are the differences in total baryon number and total entropy between the two solutions. Therefore, the terms
in Eq.~\eqref{deltaM} depending on the matter variables vanish, if the two solutions have the same entropy and baryon number [c.f. Eq.~\eqref{eq:sens}]

Also, if the two solutions are asymptotically flat, i.e.~$g_{\mu\nu} =\eta_{\mu\nu}+{\cal O}(1/r)$, 
the first term in Eq.~\eqref{deltaM} also vanishes~\cite{damour_esposito_farese}.
 This is seen by
evaluating the integral at $r\to \infty$, since the integrand decays as $1/r^3$, while $d^2 S_r\sim r^2$. 
Therefore, the mass variation only depends on the non-GR part of the action, i.e.
\begin{equation}\label{deltaMfinal}
\delta M =- \int d^2 S_i \frac{\partial {\cal L}_\phi}{\partial (\partial_i \phi_A)}\delta \phi_A
-\int d^2 S_i \frac{\partial {\cal L}_{\phi}}{\partial (\partial_i g_{\mu\nu})}\delta g_{\mu\nu}\,.
\end{equation}
This generalizes similar expressions for scalar-tensor theories~\cite{damour_esposito_farese}, and Lorentz-violating gravity~\cite{PhysRevD.89.084067,Yagi:2013qpa}.

Let us now consider an action depending
also on \textit{second} derivatives of the metric and extra gravitational degrees of freedom $\phi_A$ (e.g.~Horndeski theories):
\begin{equation}
S=\int {\cal L}(Q_A,\partial_\mu Q_A,\partial_\nu\partial_\mu Q_A) d^4 x\,.\label{2ndorder}
\end{equation}
By introducing new fields $X_{A\mu}\equiv\partial_\mu Q_A$ and enforcing this definition by Lagrangian multipliers, one obtains
\begin{equation}
S=\int \big[{\cal L}(Q_A,X_{A\mu},\partial_\nu X_{A\mu})+\lambda^{A\mu}(X_{A\mu}-\partial_\mu Q_A)\big] d^4 x\,,\label{1storder}
\end{equation}
whose variation relative to $X_{A\mu}$ yields 
\begin{equation}\label{lagrange}
\lambda^{A\mu}=
\partial_\alpha \left(\frac{\partial \cal{L}}{\partial (\partial_\alpha X_{A\mu})}\right)-\frac{\partial \cal{L}}{\partial X_{A\mu}}\,.
\end{equation}
Since Eq.~\eqref{1storder}
 is in the form given by Eq.~\eqref{generic_action}, the construction outlined above gives [using also
 Eq.~\eqref{lagrange}]
\begin{multline}\label{deltaMfinal2}
\delta M =-\int d^2 S_i \frac{\partial {\cal L}_\phi}{\partial (\partial_i \phi_A)}\delta \phi_A\\
-\int d^2 S_i \frac{\partial {\cal L}_{\phi}}{\partial (\partial_i g_{\mu\nu})}\delta g_{\mu\nu}
- \int d^2 S_i \frac{\partial {\cal L}_\phi}{\partial (\partial_i\partial_j \phi_A)}\partial_j\delta \phi_A\\
-\int d^2 S_i \frac{\partial {\cal L}_{\phi}}{\partial (\partial_i\partial_j g_{\mu\nu})}\partial_j\delta g_{\mu\nu}
+ \int d^2 S_i \partial_j\left(\frac{\partial {\cal L}_\phi}{\partial (\partial_i\partial_j \phi_A)}\right)\delta \phi_A\\
+\int d^2 S_i \partial_j \left(\frac{\partial {\cal L}_{\phi}}{\partial (\partial_i\partial_j g_{\mu\nu})}\right)\delta g_{\mu\nu}
\,.
\end{multline}

%%%%%%%%%%%%%%%%%%%%%%%%%%%%%
\emph{Sensitivities and gravitational radiation in SSHTs}:
SSHTs are described
by the Lagrangian for the ``galileon'' scalar $\phi$~\cite{Horndeski:1974wa,Deffayet:2009mn,Deffayet:2011gz}:
\begin{eqnarray}
\label{lagr}
 {\cal L}_{\phi}&&=\frac{\sqrt{-g}}{16 \pi G}\Big\{ K(X) -G_3(X)\Box\phi + G_{4}(X)R\nonumber\\&&
+G_{4X}\left[
\left(\Box\phi\right)^2-\left(\nabla_\mu\nabla_\nu\phi\right)^2
\right]\nonumber
\\
&&+G_5(X) G_{\mu\nu}\nabla^\mu\nabla^\nu\phi
-\frac{G_{5X}}{6}\Bigl[
\left(\Box\phi\right)^3
\nonumber\\&&
-3\left(\Box\phi\right)\left(\nabla_\mu\nabla_\nu\phi\right)^2
+2\left(\nabla_\mu\nabla_\nu\phi\right)^3
\Bigr]+\chi \phi {\cal G}\Big\},\,\,\,\,\,\,
\end{eqnarray}
where $\nabla$, $R$ and $G_{\mu\nu}$ are the Levi-Civita connection, Ricci scalar and Einstein tensor, 
$K$, $G_3$, $G_4$, and $G_5$ are arbitrary functions of $X\equiv-\nabla_\mu\phi \nabla^\mu\phi/2$,  $G_{iX}\equiv \partial G_i/\partial X$, 
$\Box \equiv \nabla^\mu\nabla_\mu$, $\left(\nabla_\mu\nabla_\nu\phi\right)^2  \equiv \nabla_\mu\nabla^\nu \phi\nabla_\nu\nabla^\mu\phi$,  
$\left(\nabla_\mu\nabla_\nu\phi\right)^3 \equiv \nabla_\mu\nabla^\rho \phi\nabla_\rho\nabla^\nu\phi\nabla_\nu\nabla^\mu\phi$,
$\chi$ is a constant, and ${\cal G}\equiv R^{\mu\nu\lambda\kappa}R_{\mu\nu\lambda\kappa}-4 R^{\mu\nu}R_{\mu\nu}+R^2$ is
the Gauss-Bonnet scalar\footnote{The $\chi \phi {\cal G}$ term is shift-invariant 
because ${\cal G}$ is (locally) a total divergence. Also,
this term can be obtained by choosing $G_5\propto\ln\vert X\vert$~\cite{Kobayashi:2011nu}.}. 
The total action also includes 
the Einstein Lagrangian ${\cal L}_g$ and the perfect-fluid Lagrangian ${\cal L}_{m}$ described above.
Regarding the latter, one may couple the matter fields to a ``disformal'' metric $\tilde{g}_{\mu\nu}=g_{\mu\nu}+\xi \nabla_\mu \phi \nabla_\nu \phi$
($\xi$ being a constant), rather than to  $g_{\mu\nu}$ alone. If such a disformal coupling is present, however, one can adopt $\tilde{g}_{\mu\nu}$ as the metric field, which puts the matter Lagrangian in the form ${\cal L}_m(\psi_B,\partial_\mu \psi_B,\tilde{g}_{\mu\nu})$ considered above, while
the action \eqref{lagr} remains invariant up to redefinitions of the functions $K$, $G_3$, $G_4$, and $G_5$~\cite{Bettoni:2013diz}. Thus, our results
also apply to SSHTs with a (special) disformal coupling to matter.
To allow asymptotically flat solutions (see below), we assume that $K$, $G_3$, $G_4$, and $G_5$ are analytic in X
\footnote{This excludes e.g. $K(X)\sim X^{3/2}$, which reproduces MOdified Newtonian Dynamics (MOND)~\cite{Famaey:2011kh} in the non-relativistic limit, 
and gives $\phi\sim \ln r$ near spatial infinity.}.
This implies $K(X)=X+{\cal O}(X)^2$, since a constant can be absorbed in the matter stress-energy tensor
as an effective cosmological constant, and a coefficient for the linear term can be absorbed by redefining $\phi$; 
$G_3={\cal O}(X)$ and $G_5={\cal O}(X)$, since a constant produces a total divergence\footnote{Recall the Bianchi identity $\nabla_\mu G^{\mu\nu}=0$.}; and $G_4={\cal O}(X)$, since a constant in $G_4$ can be absorbed in the metric Lagrangian  ${\cal L}_g$ by redefining the ``bare'' Newton constant $G$.

Consider now an isolated stationary star, i.e.~$\phi= {\cal O} (1/r)$, 
$g_{\mu\nu}=\eta_{\mu\nu} + {\cal O} (1/r)$, where the
shift symmetry allows setting $\phi$ asymptotically to zero. 
[In theories with a Vainshtein screening, $\phi={\cal O} (1/r)$ and $g_{\mu\nu}=\eta_{\mu\nu} + {\cal O} (1/r)$
only for $r\gg r_{\rm v}$, with $r_{\rm v}$ the Vainshtein radius within which deviations from GR are screened; we will return to this later.]
To determine the sensitivities, recall that  Eq.~\eqref{deltaMfinal2} compares two neighboring solutions. If the latter
are asymptotically flat, $G_{\mu\nu}\sim R\sim R_{\mu\nu}\sim R_{\mu\nu\alpha\beta}={\cal O}(1/r)^3$, and the differences $\delta \phi$ and $\delta g_{\mu\nu}$ between them
 scale as 
 $\delta \phi={\cal O} (1/r)$ and $\delta g_{\mu\nu}={\cal O}(1/r)$.
With these asymptotics, all the surface integrals  in  Eq.~\eqref{deltaMfinal2} vanish when evaluated at $r\to\infty$, 
thus yielding zero sensitivities.
For example, by considering the contribution of $K$ to the first term in Eq.~\eqref{deltaMfinal2}, we find
\begin{equation}\label{sens_expl}
\delta M \sim  \int d^2 S_i \frac{\partial {\cal L}_\phi}{\partial (\partial_i \phi)}\delta \phi \sim r^2 |\nabla \phi| \frac{1}{r} \sim \frac{1}{r}\,,
\end{equation}
and this surface integral vanishes when evaluated at $r\to\infty$. 
Similar calculations show that all terms in  Eq.~\eqref{deltaMfinal2} vanish.

Note that it is the shift symmetry that makes the sensitivities vanish. In
generic Horndeski theories, 
two solutions (``1'' and ``2'') have in general different asymptotic values of $\phi$, i.e.~$\phi^{(1)}=\phi_{\infty}+\alpha/r+{\cal O}(1/r^2)$ ($\alpha$ being a constant)
and $\phi^{(2)}=\phi_{\infty}+\delta \phi_{\infty}+(\alpha+\delta \alpha)/r+{\cal O}(1/r^2)$.
As such, $\delta\phi=\delta \phi_{\infty}+{\cal O}(1/r)$
(while $\delta\phi={\cal O}(1/r)$ in SSHTs) and Eq.~\eqref{sens_expl}
gives $\delta M\propto \alpha \delta \phi_{\infty}$, hence 
$s_{\phi} \propto \partial M/\partial \phi_{\infty}\propto \alpha$. This is e.g.~the case for FJBD theory~\cite{Fierz:1956zz,Jordan:1959eg,Brans:1961sx} and Damour-Esposito-Far\`{e}se gravity~\cite{ST0}, where the sensitivities are proportional to the
coefficient $\alpha$ of the $1/r$-term in the scalar field's fall-off.
Therefore, these theories
predict the emission of dipolar radiation, hence they can be constrained by existing binary-pulsar data~\cite{freire}, and give testable predictions for upcoming GW interferometers~\cite{ST1,ST2,Sampson:2014qqa,Will:2004xi,Berti:2004bd}. 
Also, in general the sensitivities are not zero in the presence of
a conformal coupling between the galileon and matter, e.g.~in massive gravity. This effect
 is not considered e.g.~in Refs.~\cite{deRham:2012fw,deRham:2012fg}.

The vanishing sensitivities imply the absence of dipolar gravitational emission 
in SSHTs. 
Consider metric and scalar perturbations $h_{\mu\nu}$ and $\delta\phi$ over a Minkowski background, i.e.~$g_{\mu\nu}=\eta_{\mu\nu}+\epsilon h_{\mu\nu}+{\cal O}(\epsilon)^2$ and $\phi=\epsilon \delta\phi+{\cal O}(\epsilon)^2$,
with $\epsilon$ a perturbative parameter. To leading order in $\epsilon$, Eq.~\eqref{lagr}  coincides with the Lagrangian
of a minimally coupled scalar field\footnote{Note 
that up to boundary terms, $G_4={\cal O}(X)$ gives ${\cal O}(\epsilon)^3$-terms in the action, because
$X R+(\Box \phi)^2-(\nabla_\mu\nabla_\nu \phi)^2=G_{\mu\nu} \nabla^\mu \phi\nabla^\nu \phi\,+\,$total divergence~\cite{Kobayashi:2011nu}.}, i.e.~the leading-order field equations
for a binary become
\begin{align}
\Box_\eta \bar{h}^{\mu\nu} &= -16 \pi G T^{\mu\nu} +\epsilon\,{\cal O}(\bar{h} \delta \phi, \delta\phi^2,\bar{h}^2)\label{eqh}\,,\\
\Box_\eta \delta\phi &= {\cal O}(s^{_{(1)}}_{\phi},s^{_{(2)}}_{\phi})  +\epsilon\,{\cal O}(\bar{h} \delta \phi, \delta\phi^2,\bar{h}^2)\notag\\&\qquad\qquad\quad=\epsilon\,{\cal O}(\bar{h} \delta \phi, \delta\phi^2, \bar{h}^2)\label{eqphi} \,,
\end{align}
where $\Box_\eta=\eta^{\mu\nu}\partial_\mu\partial_\nu$, $\bar{h}^{\mu\nu} = \eta^{\mu\nu} - \sqrt{-g} g^{\mu\nu}=
h^{\mu\nu}-\frac12 \eta^{\mu\nu} h^\alpha_{\phantom{\alpha}\alpha}+{\cal O}(\epsilon)$,
$s^{_{(1)}}_{\phi}=s^{_{(2)}}_{\phi}=0$ are the sensitivities, $T^{\mu\nu}$ is the binary's stress energy tensor (which is the same as in GR and depends
on the stars' masses and velocities), and we have assumed the Lorenz gauge $\partial_\mu \bar{h}^{\mu\nu}=0$. (These equations
can also be obtained from Ref.~\cite{damour_esposito_farese}, by noting that the theories studied here 
coincide at leading order with those of Ref.~\cite{damour_esposito_farese}, when the conformal scalar-matter coupling 
is switched off, i.e.~$A(\phi)=1$ in Ref.~\cite{damour_esposito_farese}'s notation.) Equation~\eqref{eqphi} implies
that $\phi$ is not excited at leading order in $\epsilon$,
while Eq.~\eqref{eqh} matches its GR counterpart. As such, 
gravitational emission behaves as in GR at leading PN order, i.e.~monopolar and dipolar emission 
vanish -- like in GR, but unlike non-shift-symmetric scalar-tensor theories, where these emission channels are
sourced by the sensitivities~\cite{damour_esposito_farese,zaglauer} --, while the quadrupolar emission matches GR's\footnote{FJBD theory
corrects GR's quadrupole formula even for $s^{_{(1)}}_{\phi}=s^{_{(2)}}_{\phi}=0$, due to the (Einstein-frame) conformal scalar-matter coupling.
This coupling (and corresponding corrections) are not present here.}.

Replace now the leading-order solution in the non-linear terms of Eqs.~\eqref{eqh} -- \eqref{eqphi}. Since $\delta \phi=0$ at leading order, at next-to-leading order
one has
\begin{align}
\Box_\eta \bar{h}^{\mu\nu} &= -16 \pi G [ (1- \epsilon \bar{h}^\alpha{}_\alpha) T^{\mu\nu}+\epsilon \tau^{\mu\nu}] +{\cal O}(\epsilon)^2\label{eqh2}\,,\\
\Box_\eta \delta\phi &= - \epsilon \chi \delta\mathcal{G}+{\cal O}(\epsilon)^2\label{eqphi2} \,.
\end{align}
Here, $\tau^{\mu\nu}={\cal O}(\bar{h}^2)$ is GR's gravitational stress-energy pseudotensor~\cite{2000PhRvD..62l4015P}, while $\delta\mathcal{G}={\cal O}(\bar{h})^2$ is the perturbed Gauss-Bonnet invariant. 
Equations~\eqref{eqh2} -- \eqref{eqphi2} match those of Einstein-dilaton Gauss-Bonnet gravity [i.e.~$K = X$ and $G_3 = G_4 = G_5 = 0$ in Eq.~\eqref{lagr}] in the decoupling limit,
hence the leading PN order at which deviations from GR appear is the same as 
in that theory. Following Ref.~\cite{Yagi:2015oca}, 
we then conclude that non-GR effects only appear at 3PN (2PN) order in the dissipative (conservative) sector.

Our PN formalism is only valid for stellar radii $R$ much smaller than the gravitational wavelength $\lambda_{_{\rm GW}}$, so that 
the perturbations decay as $1/r$
in both the far zone (i.e.~at distances from the binary $r\gg \lambda_{_{\rm GW}}$) and near zone (i.e.~for $R\ll r\ll\lambda_{_{\rm GW}}$)~\cite{Blanchet:2013haa}.
In theories with a Vainshtein mechanism, $\phi={\cal O} (1/r)$ and $g_{\mu\nu}=\eta_{\mu\nu} + {\cal O} (1/r)$ only
for $r\gg r_{\rm v}$, i.e.~$r_{\rm v}$ is an ``effective'' stellar radius. 
Therefore, our perturbative PN approach requires $r_{\rm v}\ll \lambda_{_{\rm GW}}$,
in which case $r_{\rm v}$ only causes higher-PN order ``finite-size'' effects~\cite{Blanchet:2013haa}.
Note that for the dominant quadrupole mode, $\lambda_{_{\rm GW}}\sim 10^{9}$ km for binary pulsars and 
$\lambda_{_{\rm GW}}\sim 10^3$ km for the late inspiral of neutron-star binaries targeted by upcoming GW detectors.
Thus, although the value of $r_{\rm v}$ (and the very presence of a Vainshtein mechanism) depend on the theory,
our approach might break down, especially in the latter case. 
If the dynamics is not perturbative/PN the analysis may be conducted
case by case using a WKB approach~\cite{deRham:2012fw,deRham:2012fg}. While simple theories (e.g.~cubic galileons)
may still provide results in agreement with binary-pulsar data~\cite{deRham:2012fw} [see however also Ref.~\cite{Chu:2012kz}], in more generic theories (e.g.~ones including quartic and quintic galileons), the non-perturbative dynamics generally makes the WKB approach also 
fail unless $r_{\rm v}\ll \lambda_{_{\rm GW}}$~\cite{deRham:2012fg}. If this condition is not satisfied, Ref.~\cite{deRham:2012fg} concludes
that many multipoles radiate with the same strength. This seems difficult to reconcile with binary-pulsar observations,
which agree with GR's quadrupole formula. (Note however that unlike us, Ref.~\cite{deRham:2012fg} assumes a conformal scalar-matter coupling.)

In conclusion, GW emission from stellar binaries is only modified at high PN orders in SSHTs.
Therefore, binary-pulsar observations, which agree with GR's
quadrupole formula at percent-level and also test the 1PN conservative dynamics, are reproduced in SSHTs, provided
that the binary's dynamics is perturbative/PN~\cite{deRham:2012fg}, and that cosmological-expansion effects can be neglected~\cite{Jimenez:2015bwa,Babichev:2012re,Babichev:2013cya,Jacobson:1999vr}.
Direct detection of GWs from neutron-star binaries, however, may still provide prospects for testing these theories.
Also, our main result Eq.~\eqref{deltaMfinal2} applies to generic gravitational theories,
but only to stars and not black holes (where additional surface integrals at the horizon might be present).
In the special case of SSHTs, black holes are the same as in GR if $\chi=0$ in Eq.~\eqref{lagr}~\cite{Hui:2012qt},
but possess scalar hairs if $\chi\neq0$~\cite{Sotiriou:2013qea,Sotiriou:2014pfa,Maselli:2015yva}. 
We thus expect dipolar emission from binaries involving black holes  
in SSHTs with $\chi\neq0$~\cite{kent-LMXB,Yagi:2015oca}.

\begin{acknowledgments}
\emph{Acknowledgments}.--- We acknowledge support from
 the European Union's Seventh Framework Programme (FP7/PEOPLE-2011-CIG)
through the  Marie Curie Career Integration Grant GALFORMBHS PCIG11-GA-2012-321608,  and from the H2020-MSCA-RISE-2015 Grant No. StronGrHEP-690904 (to E.B.);
NSF CAREER Award PHY-1250636 and JSPS Postdoctoral Fellowships for Research Abroad (to K.Y.).
K.Y. also thanks the Institut d'Astrophysique de Paris for the hospitality during a visit in which
the early idea for this work was conceived. We are especially indebted to Gilles  Esposito-Far\`{e}se for providing, as usual, deep, stimulating
and enlightening comments. We also thank Luis Lehner, Leo Stein and Clifford Will for reading
a preliminary version of this manuscript and providing insightful feedback, as well as Guillaume Faye for useful discussions about post-Newtonian theory.
\end{acknowledgments}

\bibliography{master}

\begin{thebibliography}{56}
\expandafter\ifx\csname natexlab\endcsname\relax\def\natexlab#1{#1}\fi
\expandafter\ifx\csname bibnamefont\endcsname\relax
  \def\bibnamefont#1{#1}\fi
\expandafter\ifx\csname bibfnamefont\endcsname\relax
  \def\bibfnamefont#1{#1}\fi
\expandafter\ifx\csname citenamefont\endcsname\relax
  \def\citenamefont#1{#1}\fi
\expandafter\ifx\csname url\endcsname\relax
  \def\url#1{\texttt{#1}}\fi
\expandafter\ifx\csname urlprefix\endcsname\relax\def\urlprefix{URL }\fi
\providecommand{\bibinfo}[2]{#2}
\providecommand{\eprint}[2][]{\url{#2}}

\bibitem[{\citenamefont{Damour and Taylor}(1992)}]{damour-taylor}
\bibinfo{author}{\bibfnamefont{T.}~\bibnamefont{Damour}} \bibnamefont{and}
  \bibinfo{author}{\bibfnamefont{J.~H.} \bibnamefont{Taylor}},
  \bibinfo{journal}{Phys. Rev. D} \textbf{\bibinfo{volume}{45}},
  \bibinfo{pages}{1840} (\bibinfo{year}{1992}).

\bibitem[{lig()}]{ligo}
\emph{\bibinfo{title}{{LIGO}}}, \bibinfo{note}{{\tt www.ligo.caltech.edu}}.

\bibitem[{vir()}]{virgo}
\emph{\bibinfo{title}{{VIRGO}}}, \bibinfo{note}{{\tt www.virgo.infn.it}}.

\bibitem[{kag()}]{kagra}
\emph{\bibinfo{title}{Kagra}}, \bibinfo{note}{{\tt
  http://gwcenter.icrr.u-tokyo.ac.jp/en/}}.

\bibitem[{\citenamefont{Fierz}(1956)}]{Fierz:1956zz}
\bibinfo{author}{\bibfnamefont{M.}~\bibnamefont{Fierz}},
  \bibinfo{journal}{Helv. Phys. Acta} \textbf{\bibinfo{volume}{29}},
  \bibinfo{pages}{128} (\bibinfo{year}{1956}).

\bibitem[{\citenamefont{Jordan}(1959)}]{Jordan:1959eg}
\bibinfo{author}{\bibfnamefont{P.}~\bibnamefont{Jordan}}, \bibinfo{journal}{Z.
  Phys.} \textbf{\bibinfo{volume}{157}}, \bibinfo{pages}{112}
  (\bibinfo{year}{1959}).

\bibitem[{\citenamefont{Brans and Dicke}(1961)}]{Brans:1961sx}
\bibinfo{author}{\bibfnamefont{C.}~\bibnamefont{Brans}} \bibnamefont{and}
  \bibinfo{author}{\bibfnamefont{R.~H.} \bibnamefont{Dicke}},
  \bibinfo{journal}{Phys. Rev.} \textbf{\bibinfo{volume}{124}},
  \bibinfo{pages}{925} (\bibinfo{year}{1961}).

\bibitem[{\citenamefont{Berti et~al.}(2015)}]{Berti:2015itd}
\bibinfo{author}{\bibfnamefont{E.}~\bibnamefont{Berti}} \bibnamefont{et~al.}
  (\bibinfo{year}{2015}), \eprint{1501.07274}.

\bibitem[{\citenamefont{{Eardley}}(1975)}]{eardley}
\bibinfo{author}{\bibfnamefont{D.~M.} \bibnamefont{{Eardley}}},
  \bibinfo{journal}{Astrophys. J. Lett.} \textbf{\bibinfo{volume}{196}},
  \bibinfo{pages}{L59} (\bibinfo{year}{1975}).

\bibitem[{\citenamefont{Yagi et~al.}(2014{\natexlab{a}})\citenamefont{Yagi,
  Blas, Barausse, and Yunes}}]{PhysRevD.89.084067}
\bibinfo{author}{\bibfnamefont{K.}~\bibnamefont{Yagi}},
  \bibinfo{author}{\bibfnamefont{D.}~\bibnamefont{Blas}},
  \bibinfo{author}{\bibfnamefont{E.}~\bibnamefont{Barausse}}, \bibnamefont{and}
  \bibinfo{author}{\bibfnamefont{N.}~\bibnamefont{Yunes}},
  \bibinfo{journal}{Phys. Rev. D} \textbf{\bibinfo{volume}{89}},
  \bibinfo{pages}{084067} (\bibinfo{year}{2014}{\natexlab{a}}).

\bibitem[{\citenamefont{Yagi et~al.}(2014{\natexlab{b}})\citenamefont{Yagi,
  Blas, Yunes, and Barausse}}]{Yagi:2013qpa}
\bibinfo{author}{\bibfnamefont{K.}~\bibnamefont{Yagi}},
  \bibinfo{author}{\bibfnamefont{D.}~\bibnamefont{Blas}},
  \bibinfo{author}{\bibfnamefont{N.}~\bibnamefont{Yunes}}, \bibnamefont{and}
  \bibinfo{author}{\bibfnamefont{E.}~\bibnamefont{Barausse}},
  \bibinfo{journal}{Phys. Rev. Lett.} \textbf{\bibinfo{volume}{112}},
  \bibinfo{pages}{161101} (\bibinfo{year}{2014}{\natexlab{b}}).

\bibitem[{\citenamefont{Freire et~al.}(2012)\citenamefont{Freire, Wex,
  Esposito-Farese, Verbiest, Bailes et~al.}}]{freire}
\bibinfo{author}{\bibfnamefont{P.~C.} \bibnamefont{Freire}},
  \bibinfo{author}{\bibfnamefont{N.}~\bibnamefont{Wex}},
  \bibinfo{author}{\bibfnamefont{G.}~\bibnamefont{Esposito-Farese}},
  \bibinfo{author}{\bibfnamefont{J.~P.} \bibnamefont{Verbiest}},
  \bibinfo{author}{\bibfnamefont{M.}~\bibnamefont{Bailes}},
  \bibnamefont{et~al.}, \bibinfo{journal}{Mon. Not. Roy. Astron. Soc.}
  \textbf{\bibinfo{volume}{423}}, \bibinfo{pages}{3328} (\bibinfo{year}{2012}).

\bibitem[{\citenamefont{Wex}(2014)}]{Wex:2014nva}
\bibinfo{author}{\bibfnamefont{N.}~\bibnamefont{Wex}} (\bibinfo{year}{2014}),
  \eprint{1402.5594}.

\bibitem[{\citenamefont{Barausse et~al.}(2013)\citenamefont{Barausse,
  Palenzuela, Ponce, and Lehner}}]{ST1}
\bibinfo{author}{\bibfnamefont{E.}~\bibnamefont{Barausse}},
  \bibinfo{author}{\bibfnamefont{C.}~\bibnamefont{Palenzuela}},
  \bibinfo{author}{\bibfnamefont{M.}~\bibnamefont{Ponce}}, \bibnamefont{and}
  \bibinfo{author}{\bibfnamefont{L.}~\bibnamefont{Lehner}},
  \bibinfo{journal}{Phys. Rev. D} \textbf{\bibinfo{volume}{87}},
  \bibinfo{pages}{081506} (\bibinfo{year}{2013}).

\bibitem[{\citenamefont{Palenzuela et~al.}(2014)\citenamefont{Palenzuela,
  Barausse, Ponce, and Lehner}}]{ST2}
\bibinfo{author}{\bibfnamefont{C.}~\bibnamefont{Palenzuela}},
  \bibinfo{author}{\bibfnamefont{E.}~\bibnamefont{Barausse}},
  \bibinfo{author}{\bibfnamefont{M.}~\bibnamefont{Ponce}}, \bibnamefont{and}
  \bibinfo{author}{\bibfnamefont{L.}~\bibnamefont{Lehner}},
  \bibinfo{journal}{Phys. Rev. D} \textbf{\bibinfo{volume}{89}},
  \bibinfo{pages}{044024} (\bibinfo{year}{2014}).

\bibitem[{\citenamefont{Sampson et~al.}(2014)\citenamefont{Sampson, Yunes,
  Cornish, Ponce, Barausse, Klein, Palenzuela, and Lehner}}]{Sampson:2014qqa}
\bibinfo{author}{\bibfnamefont{L.}~\bibnamefont{Sampson}},
  \bibinfo{author}{\bibfnamefont{N.}~\bibnamefont{Yunes}},
  \bibinfo{author}{\bibfnamefont{N.}~\bibnamefont{Cornish}},
  \bibinfo{author}{\bibfnamefont{M.}~\bibnamefont{Ponce}},
  \bibinfo{author}{\bibfnamefont{E.}~\bibnamefont{Barausse}},
  \bibinfo{author}{\bibfnamefont{A.}~\bibnamefont{Klein}},
  \bibinfo{author}{\bibfnamefont{C.}~\bibnamefont{Palenzuela}},
  \bibnamefont{and} \bibinfo{author}{\bibfnamefont{L.}~\bibnamefont{Lehner}},
  \bibinfo{journal}{Phys. Rev. D} \textbf{\bibinfo{volume}{90}},
  \bibinfo{pages}{124091} (\bibinfo{year}{2014}).

\bibitem[{\citenamefont{{Damour} and
  {Esposito-Farese}}(1992)}]{damour_esposito_farese}
\bibinfo{author}{\bibfnamefont{T.}~\bibnamefont{{Damour}}} \bibnamefont{and}
  \bibinfo{author}{\bibfnamefont{G.}~\bibnamefont{{Esposito-Farese}}},
  \bibinfo{journal}{Classical and Quantum Gravity}
  \textbf{\bibinfo{volume}{9}}, \bibinfo{pages}{2093} (\bibinfo{year}{1992}).

\bibitem[{\citenamefont{{Damour} and {Esposito-Farese}}(1993)}]{ST0}
\bibinfo{author}{\bibfnamefont{T.}~\bibnamefont{{Damour}}} \bibnamefont{and}
  \bibinfo{author}{\bibfnamefont{G.}~\bibnamefont{{Esposito-Farese}}},
  \bibinfo{journal}{Phys. Rev. Lett.} \textbf{\bibinfo{volume}{70}},
  \bibinfo{pages}{2220} (\bibinfo{year}{1993}).

\bibitem[{\citenamefont{Horndeski}(1974)}]{Horndeski:1974wa}
\bibinfo{author}{\bibfnamefont{G.~W.} \bibnamefont{Horndeski}},
  \bibinfo{journal}{Int. J. Theor. Phys.} \textbf{\bibinfo{volume}{10}},
  \bibinfo{pages}{363} (\bibinfo{year}{1974}).

\bibitem[{\citenamefont{Deffayet et~al.}(2009)\citenamefont{Deffayet, Deser,
  and Esposito-Farese}}]{Deffayet:2009mn}
\bibinfo{author}{\bibfnamefont{C.}~\bibnamefont{Deffayet}},
  \bibinfo{author}{\bibfnamefont{S.}~\bibnamefont{Deser}}, \bibnamefont{and}
  \bibinfo{author}{\bibfnamefont{G.}~\bibnamefont{Esposito-Farese}},
  \bibinfo{journal}{Phys. Rev. D} \textbf{\bibinfo{volume}{80}},
  \bibinfo{pages}{064015} (\bibinfo{year}{2009}).

\bibitem[{\citenamefont{Deffayet et~al.}(2011)\citenamefont{Deffayet, Gao,
  Steer, and Zahariade}}]{Deffayet:2011gz}
\bibinfo{author}{\bibfnamefont{C.}~\bibnamefont{Deffayet}},
  \bibinfo{author}{\bibfnamefont{X.}~\bibnamefont{Gao}},
  \bibinfo{author}{\bibfnamefont{D.~A.} \bibnamefont{Steer}}, \bibnamefont{and}
  \bibinfo{author}{\bibfnamefont{G.}~\bibnamefont{Zahariade}},
  \bibinfo{journal}{Phys. Rev. D} \textbf{\bibinfo{volume}{84}},
  \bibinfo{pages}{064039} (\bibinfo{year}{2011}).

\bibitem[{\citenamefont{Vainshtein}(1972)}]{Vainshtein:1972sx}
\bibinfo{author}{\bibfnamefont{A.~I.} \bibnamefont{Vainshtein}},
  \bibinfo{journal}{Phys. Lett. B} \textbf{\bibinfo{volume}{39}},
  \bibinfo{pages}{393} (\bibinfo{year}{1972}).

\bibitem[{\citenamefont{Deffayet et~al.}(2002)\citenamefont{Deffayet, Dvali,
  Gabadadze, and Vainshtein}}]{Deffayet:2001uk}
\bibinfo{author}{\bibfnamefont{C.}~\bibnamefont{Deffayet}},
  \bibinfo{author}{\bibfnamefont{G.~R.} \bibnamefont{Dvali}},
  \bibinfo{author}{\bibfnamefont{G.}~\bibnamefont{Gabadadze}},
  \bibnamefont{and} \bibinfo{author}{\bibfnamefont{A.~I.}
  \bibnamefont{Vainshtein}}, \bibinfo{journal}{Phys. Rev. D}
  \textbf{\bibinfo{volume}{65}}, \bibinfo{pages}{044026}
  (\bibinfo{year}{2002}).

\bibitem[{\citenamefont{Babichev and Deffayet}(2013)}]{Babichev:2013usa}
\bibinfo{author}{\bibfnamefont{E.}~\bibnamefont{Babichev}} \bibnamefont{and}
  \bibinfo{author}{\bibfnamefont{C.}~\bibnamefont{Deffayet}},
  \bibinfo{journal}{Class. Quant. Grav.} \textbf{\bibinfo{volume}{30}},
  \bibinfo{pages}{184001} (\bibinfo{year}{2013}).

\bibitem[{\citenamefont{Winther and Ferreira}(2015)}]{Winther:2015pta}
\bibinfo{author}{\bibfnamefont{H.~A.} \bibnamefont{Winther}} \bibnamefont{and}
  \bibinfo{author}{\bibfnamefont{P.~G.} \bibnamefont{Ferreira}},
  \bibinfo{journal}{Phys. Rev. D} \textbf{\bibinfo{volume}{92}},
  \bibinfo{pages}{064005} (\bibinfo{year}{2015}).

\bibitem[{\citenamefont{de~Rham and Gabadadze}(2010)}]{deRham:2010ik}
\bibinfo{author}{\bibfnamefont{C.}~\bibnamefont{de~Rham}} \bibnamefont{and}
  \bibinfo{author}{\bibfnamefont{G.}~\bibnamefont{Gabadadze}},
  \bibinfo{journal}{Phys. Rev. D} \textbf{\bibinfo{volume}{82}},
  \bibinfo{pages}{044020} (\bibinfo{year}{2010}).

\bibitem[{\citenamefont{de~Rham et~al.}(2011)\citenamefont{de~Rham, Gabadadze,
  and Tolley}}]{deRham:2010kj}
\bibinfo{author}{\bibfnamefont{C.}~\bibnamefont{de~Rham}},
  \bibinfo{author}{\bibfnamefont{G.}~\bibnamefont{Gabadadze}},
  \bibnamefont{and} \bibinfo{author}{\bibfnamefont{A.~J.}
  \bibnamefont{Tolley}}, \bibinfo{journal}{Phys. Rev. Lett.}
  \textbf{\bibinfo{volume}{106}}, \bibinfo{pages}{231101}
  (\bibinfo{year}{2011}).

\bibitem[{\citenamefont{Yagi et~al.}(2012)\citenamefont{Yagi, Stein, Yunes, and
  Tanaka}}]{Yagi:2011xp}
\bibinfo{author}{\bibfnamefont{K.}~\bibnamefont{Yagi}},
  \bibinfo{author}{\bibfnamefont{L.~C.} \bibnamefont{Stein}},
  \bibinfo{author}{\bibfnamefont{N.}~\bibnamefont{Yunes}}, \bibnamefont{and}
  \bibinfo{author}{\bibfnamefont{T.}~\bibnamefont{Tanaka}},
  \bibinfo{journal}{Phys. Rev. D} \textbf{\bibinfo{volume}{85}},
  \bibinfo{pages}{064022} (\bibinfo{year}{2012}).

\bibitem[{\citenamefont{Jiménez et~al.}(2015)\citenamefont{Jiménez, Piazza,
  and Velten}}]{Jimenez:2015bwa}
\bibinfo{author}{\bibfnamefont{J.~B.} \bibnamefont{Jiménez}},
  \bibinfo{author}{\bibfnamefont{F.}~\bibnamefont{Piazza}}, \bibnamefont{and}
  \bibinfo{author}{\bibfnamefont{H.}~\bibnamefont{Velten}}
  (\bibinfo{year}{2015}), \eprint{1507.05047}.

\bibitem[{\citenamefont{Babichev and Esposito-Farèse}(2013)}]{Babichev:2012re}
\bibinfo{author}{\bibfnamefont{E.}~\bibnamefont{Babichev}} \bibnamefont{and}
  \bibinfo{author}{\bibfnamefont{G.}~\bibnamefont{Esposito-Farèse}},
  \bibinfo{journal}{Phys. Rev. D} \textbf{\bibinfo{volume}{87}},
  \bibinfo{pages}{044032} (\bibinfo{year}{2013}).

\bibitem[{\citenamefont{Babichev and Charmousis}(2014)}]{Babichev:2013cya}
\bibinfo{author}{\bibfnamefont{E.}~\bibnamefont{Babichev}} \bibnamefont{and}
  \bibinfo{author}{\bibfnamefont{C.}~\bibnamefont{Charmousis}},
  \bibinfo{journal}{JHEP} \textbf{\bibinfo{volume}{08}}, \bibinfo{pages}{106}
  (\bibinfo{year}{2014}).

\bibitem[{\citenamefont{Jacobson}(1999)}]{Jacobson:1999vr}
\bibinfo{author}{\bibfnamefont{T.}~\bibnamefont{Jacobson}},
  \bibinfo{journal}{Phys. Rev. Lett.} \textbf{\bibinfo{volume}{83}},
  \bibinfo{pages}{2699} (\bibinfo{year}{1999}).

\bibitem[{\citenamefont{Iyer and Wald}(1994)}]{Iyer:1994ys}
\bibinfo{author}{\bibfnamefont{V.}~\bibnamefont{Iyer}} \bibnamefont{and}
  \bibinfo{author}{\bibfnamefont{R.~M.} \bibnamefont{Wald}},
  \bibinfo{journal}{Phys. Rev. D} \textbf{\bibinfo{volume}{50}},
  \bibinfo{pages}{846} (\bibinfo{year}{1994}), \eprint{gr-qc/9403028}.

\bibitem[{\citenamefont{{Landau} and {Lifshitz}}(1975)}]{landau}
\bibinfo{author}{\bibfnamefont{L.~D.} \bibnamefont{{Landau}}} \bibnamefont{and}
  \bibinfo{author}{\bibfnamefont{E.~M.} \bibnamefont{{Lifshitz}}},
  \emph{\bibinfo{title}{{The classical theory of fields}}}
  (\bibinfo{publisher}{Pergamon Press}, \bibinfo{address}{Oxford},
  \bibinfo{year}{1975}).

\bibitem[{\citenamefont{Schutz}(1970)}]{PhysRevD.2.2762}
\bibinfo{author}{\bibfnamefont{B.~F.} \bibnamefont{Schutz}},
  \bibinfo{journal}{Phys. Rev. D} \textbf{\bibinfo{volume}{2}},
  \bibinfo{pages}{2762} (\bibinfo{year}{1970}).

\bibitem[{\citenamefont{Brown}(1993)}]{Brown:1992kc}
\bibinfo{author}{\bibfnamefont{J.~D.} \bibnamefont{Brown}},
  \bibinfo{journal}{Class. Quant. Grav.} \textbf{\bibinfo{volume}{10}},
  \bibinfo{pages}{1579} (\bibinfo{year}{1993}).

\bibitem[{\citenamefont{Misner et~al.}(1973)\citenamefont{Misner, Thorne, and
  Wheeler}}]{MTW}
\bibinfo{author}{\bibfnamefont{C.~W.} \bibnamefont{Misner}},
  \bibinfo{author}{\bibfnamefont{K.}~\bibnamefont{Thorne}}, \bibnamefont{and}
  \bibinfo{author}{\bibfnamefont{J.~A.} \bibnamefont{Wheeler}},
  \emph{\bibinfo{title}{Gravitation}} (\bibinfo{publisher}{W. H. Freeman \&
  Co.}, \bibinfo{address}{San Francisco}, \bibinfo{year}{1973}).

\bibitem[{\citenamefont{{Rezzolla} and {Zanotti}}(2013)}]{rezzolla_book}
\bibinfo{author}{\bibfnamefont{L.}~\bibnamefont{{Rezzolla}}} \bibnamefont{and}
  \bibinfo{author}{\bibfnamefont{O.}~\bibnamefont{{Zanotti}}},
  \emph{\bibinfo{title}{{Relativistic Hydrodynamics}}} (\bibinfo{year}{2013}).

\bibitem[{\citenamefont{Gourgoulhon}(2006)}]{eric}
\bibinfo{author}{\bibfnamefont{E.}~\bibnamefont{Gourgoulhon}},
  \bibinfo{journal}{EAS Publ. Ser.} \textbf{\bibinfo{volume}{21}},
  \bibinfo{pages}{43} (\bibinfo{year}{2006}), \eprint{gr-qc/0603009}.

\bibitem[{\citenamefont{Kobayashi et~al.}(2011)\citenamefont{Kobayashi,
  Yamaguchi, and Yokoyama}}]{Kobayashi:2011nu}
\bibinfo{author}{\bibfnamefont{T.}~\bibnamefont{Kobayashi}},
  \bibinfo{author}{\bibfnamefont{M.}~\bibnamefont{Yamaguchi}},
  \bibnamefont{and} \bibinfo{author}{\bibfnamefont{J.}~\bibnamefont{Yokoyama}},
  \bibinfo{journal}{Prog. Theor. Phys.} \textbf{\bibinfo{volume}{126}},
  \bibinfo{pages}{511} (\bibinfo{year}{2011}).

\bibitem[{\citenamefont{Bettoni and Liberati}(2013)}]{Bettoni:2013diz}
\bibinfo{author}{\bibfnamefont{D.}~\bibnamefont{Bettoni}} \bibnamefont{and}
  \bibinfo{author}{\bibfnamefont{S.}~\bibnamefont{Liberati}},
  \bibinfo{journal}{Phys. Rev. D} \textbf{\bibinfo{volume}{88}},
  \bibinfo{pages}{084020} (\bibinfo{year}{2013}).

\bibitem[{\citenamefont{Famaey and McGaugh}(2012)}]{Famaey:2011kh}
\bibinfo{author}{\bibfnamefont{B.}~\bibnamefont{Famaey}} \bibnamefont{and}
  \bibinfo{author}{\bibfnamefont{S.}~\bibnamefont{McGaugh}},
  \bibinfo{journal}{Living Rev. Rel.} \textbf{\bibinfo{volume}{15}},
  \bibinfo{pages}{10} (\bibinfo{year}{2012}).

\bibitem[{\citenamefont{Will and Yunes}(2004)}]{Will:2004xi}
\bibinfo{author}{\bibfnamefont{C.~M.} \bibnamefont{Will}} \bibnamefont{and}
  \bibinfo{author}{\bibfnamefont{N.}~\bibnamefont{Yunes}},
  \bibinfo{journal}{Class. Quant. Grav.} \textbf{\bibinfo{volume}{21}},
  \bibinfo{pages}{4367} (\bibinfo{year}{2004}).

\bibitem[{\citenamefont{Berti et~al.}(2005)\citenamefont{Berti, Buonanno, and
  Will}}]{Berti:2004bd}
\bibinfo{author}{\bibfnamefont{E.}~\bibnamefont{Berti}},
  \bibinfo{author}{\bibfnamefont{A.}~\bibnamefont{Buonanno}}, \bibnamefont{and}
  \bibinfo{author}{\bibfnamefont{C.~M.} \bibnamefont{Will}},
  \bibinfo{journal}{Phys. Rev. D} \textbf{\bibinfo{volume}{71}},
  \bibinfo{pages}{084025} (\bibinfo{year}{2005}).

\bibitem[{\citenamefont{de~Rham
  et~al.}(2013{\natexlab{a}})\citenamefont{de~Rham, Tolley, and
  Wesley}}]{deRham:2012fw}
\bibinfo{author}{\bibfnamefont{C.}~\bibnamefont{de~Rham}},
  \bibinfo{author}{\bibfnamefont{A.~J.} \bibnamefont{Tolley}},
  \bibnamefont{and} \bibinfo{author}{\bibfnamefont{D.~H.}
  \bibnamefont{Wesley}}, \bibinfo{journal}{Phys. Rev. D}
  \textbf{\bibinfo{volume}{87}}, \bibinfo{pages}{044025}
  (\bibinfo{year}{2013}{\natexlab{a}}).

\bibitem[{\citenamefont{de~Rham
  et~al.}(2013{\natexlab{b}})\citenamefont{de~Rham, Matas, and
  Tolley}}]{deRham:2012fg}
\bibinfo{author}{\bibfnamefont{C.}~\bibnamefont{de~Rham}},
  \bibinfo{author}{\bibfnamefont{A.}~\bibnamefont{Matas}}, \bibnamefont{and}
  \bibinfo{author}{\bibfnamefont{A.~J.} \bibnamefont{Tolley}},
  \bibinfo{journal}{Phys. Rev. D} \textbf{\bibinfo{volume}{87}},
  \bibinfo{pages}{064024} (\bibinfo{year}{2013}{\natexlab{b}}).

\bibitem[{\citenamefont{Will and Zaglauer}(1989)}]{zaglauer}
\bibinfo{author}{\bibfnamefont{C.~M.} \bibnamefont{Will}} \bibnamefont{and}
  \bibinfo{author}{\bibfnamefont{H.~W.} \bibnamefont{Zaglauer}},
  \bibinfo{journal}{Astrophys. J.} \textbf{\bibinfo{volume}{346}},
  \bibinfo{pages}{366} (\bibinfo{year}{1989}).

\bibitem[{\citenamefont{{Pati} and {Will}}(2000)}]{2000PhRvD..62l4015P}
\bibinfo{author}{\bibfnamefont{M.~E.} \bibnamefont{{Pati}}} \bibnamefont{and}
  \bibinfo{author}{\bibfnamefont{C.~M.} \bibnamefont{{Will}}},
  \bibinfo{journal}{Phys. Rev. D} \textbf{\bibinfo{volume}{62}},
  \bibinfo{pages}{124015} (\bibinfo{year}{2000}).

\bibitem[{\citenamefont{Yagi et~al.}(2015)\citenamefont{Yagi, Stein, and
  Yunes}}]{Yagi:2015oca}
\bibinfo{author}{\bibfnamefont{K.}~\bibnamefont{Yagi}},
  \bibinfo{author}{\bibfnamefont{L.~C.} \bibnamefont{Stein}}, \bibnamefont{and}
  \bibinfo{author}{\bibfnamefont{N.}~\bibnamefont{Yunes}}
  (\bibinfo{year}{2015}), \eprint{1510.02152}.

\bibitem[{\citenamefont{Blanchet}(2014)}]{Blanchet:2013haa}
\bibinfo{author}{\bibfnamefont{L.}~\bibnamefont{Blanchet}},
  \bibinfo{journal}{Living Rev. Rel.} \textbf{\bibinfo{volume}{17}},
  \bibinfo{pages}{2} (\bibinfo{year}{2014}).

\bibitem[{\citenamefont{Chu and Trodden}(2013)}]{Chu:2012kz}
\bibinfo{author}{\bibfnamefont{Y.-Z.} \bibnamefont{Chu}} \bibnamefont{and}
  \bibinfo{author}{\bibfnamefont{M.}~\bibnamefont{Trodden}},
  \bibinfo{journal}{Phys. Rev. D} \textbf{\bibinfo{volume}{87}},
  \bibinfo{pages}{024011} (\bibinfo{year}{2013}).

\bibitem[{\citenamefont{Hui and Nicolis}(2013)}]{Hui:2012qt}
\bibinfo{author}{\bibfnamefont{L.}~\bibnamefont{Hui}} \bibnamefont{and}
  \bibinfo{author}{\bibfnamefont{A.}~\bibnamefont{Nicolis}},
  \bibinfo{journal}{Phys. Rev. Lett.} \textbf{\bibinfo{volume}{110}},
  \bibinfo{pages}{241104} (\bibinfo{year}{2013}).

\bibitem[{\citenamefont{Sotiriou and
  Zhou}(2014{\natexlab{a}})}]{Sotiriou:2013qea}
\bibinfo{author}{\bibfnamefont{T.~P.} \bibnamefont{Sotiriou}} \bibnamefont{and}
  \bibinfo{author}{\bibfnamefont{S.-Y.} \bibnamefont{Zhou}},
  \bibinfo{journal}{Phys. Rev. Lett.} \textbf{\bibinfo{volume}{112}},
  \bibinfo{pages}{251102} (\bibinfo{year}{2014}{\natexlab{a}}).

\bibitem[{\citenamefont{Sotiriou and
  Zhou}(2014{\natexlab{b}})}]{Sotiriou:2014pfa}
\bibinfo{author}{\bibfnamefont{T.~P.} \bibnamefont{Sotiriou}} \bibnamefont{and}
  \bibinfo{author}{\bibfnamefont{S.-Y.} \bibnamefont{Zhou}},
  \bibinfo{journal}{Phys. Rev. D} \textbf{\bibinfo{volume}{90}},
  \bibinfo{pages}{124063} (\bibinfo{year}{2014}{\natexlab{b}}).

\bibitem[{\citenamefont{Maselli et~al.}(2015)\citenamefont{Maselli, Silva,
  Minamitsuji, and Berti}}]{Maselli:2015yva}
\bibinfo{author}{\bibfnamefont{A.}~\bibnamefont{Maselli}},
  \bibinfo{author}{\bibfnamefont{H.~O.} \bibnamefont{Silva}},
  \bibinfo{author}{\bibfnamefont{M.}~\bibnamefont{Minamitsuji}},
  \bibnamefont{and} \bibinfo{author}{\bibfnamefont{E.}~\bibnamefont{Berti}}
  (\bibinfo{year}{2015}), \eprint{1508.03044}.

\bibitem[{\citenamefont{Yagi}(2012)}]{kent-LMXB}
\bibinfo{author}{\bibfnamefont{K.}~\bibnamefont{Yagi}}, \bibinfo{journal}{Phys.
  Rev. D} \textbf{\bibinfo{volume}{86}}, \bibinfo{pages}{081504}
  (\bibinfo{year}{2012}).

\end{thebibliography}

\end{document}